\documentclass[11pt]{article}

\usepackage[preprint]{acl}

\usepackage{times}
\usepackage{latexsym}
\usepackage{amsmath}
\usepackage{graphicx}
\usepackage{amssymb}
\usepackage{booktabs}
\usepackage{multirow}
\usepackage{colortbl}
\usepackage{enumitem}
\usepackage[T1]{fontenc}

\usepackage[utf8]{inputenc}

\usepackage{microtype}

\usepackage{inconsolata}

\usepackage{graphicx}
\usepackage{xcolor} 

\newcommand\blfootnote[1]{%
  \begingroup
  \renewcommand\thefootnote{}\footnote{#1}%
  \addtocounter{footnote}{-1}%
  \endgroup
}

\title{MimicLM: Zero-Shot Voice Imitation through Autoregressive Modeling of Pseudo-Parallel Speech Corpora}

\author{
  \textbf{Tao Feng\textsuperscript{1}},
  \textbf{Yuxiang Wang\textsuperscript{2}},
  \textbf{Yuancheng Wang\textsuperscript{2}},
  \textbf{Xueyao Zhang\textsuperscript{2}},
  \textbf{Dekun Chen\textsuperscript{2}},
\\
  \textbf{Chaoren Wang\textsuperscript{2}},
  \textbf{Xun Guan\textsuperscript{1}},
  \textbf{Zhizheng Wu\textsuperscript{2$\dag$}}
\\
\\
  \textsuperscript{1}Tsinghua University,
  \textsuperscript{2}The Chinese University of Hong Kong, Shenzhen
}

\begin{document}
\maketitle

\blfootnote{$\dag$ Corresponding author.}
\blfootnote{Resources available at \url{https://fff-ttt.github.io/MimicLM_demo/}}

\begin{abstract}
Voice imitation aims to transform \textit{source} speech to match a \textit{reference} speaker's timbre and speaking style while preserving linguistic content. A straightforward approach is to train on triplets of \textit{(source, reference, target)}, where \textit{source} and \textit{target} share the same content but \textit{target} matches the \textit{reference}'s voice characteristics, yet such data is extremely scarce. Existing approaches either employ carefully designed disentanglement architectures to bypass this data scarcity or leverage external systems to synthesize pseudo-parallel training data. However, the former requires intricate model design, and the latter faces a quality ceiling when synthetic speech is used as training \textit{targets}. To address these limitations, we propose MimicLM, which takes a novel approach by using synthetic speech as training \textit{sources} while retaining real recordings as \textit{targets}. This design enables the model to learn directly from real speech distributions, breaking the synthetic quality ceiling. Building on this data construction approach, we incorporate interleaved text-audio modeling to guide the generation of content-accurate speech and apply post-training with preference alignment to mitigate the inherent distributional mismatch when training on synthetic data. Experiments demonstrate that MimicLM achieves superior voice imitation quality with a simple yet effective architecture, significantly outperforming existing methods in naturalness while maintaining competitive similarity scores across speaker identity, accent, and emotion dimensions.

\end{abstract}

\section{Introduction}
Voice imitation aims to transform \textit{source} speech to match a \textit{reference} speaker's voice characteristics while preserving the linguistic content.
Unlike voice conversion that focuses solely on timbre transfer~\cite{sisman2020overview}, voice imitation additionally captures the complete speaking style including prosodic and expressive patterns~\cite{liu2020end, zhou2022emotional, zhang2025vevo}.

A straightforward approach is to train on triplets (\textit{source}, \textit{reference}, \textit{target}), where \textit{source} and \textit{target} share the same linguistic content but differ in speaker identity, while \textit{reference} provides the target speaker's voice characteristics.
However, such triplet data is extremely scarce in real-world speech corpora~\cite{yoshino2016parallel}, and manual collection at scale remains prohibitively expensive.

Existing methods address this parallel data bottleneck through two primary strategies. 
The first approach bypasses the need for parallel data by explicitly disentangling content, timbre, and prosody through specialized architectural components~\cite{qian2019autovc, qian2020unsupervised, ju2024naturalspeech, zhang2025vevo}. 
While effective, these methods require multi-stage training with carefully balanced objectives and complex inference pipelines involving multiple learned modules~\cite{ju2024naturalspeech, lajszczak2024base}. 
The second approach leverages external zero-shot text-to-speech (TTS)~\cite{wang2023neural, anastassiou2024seed, du2024cosyvoice, wang2024maskgct} or voice conversion (VC)~\cite{qin2023openvoice} systems to construct pseudo-parallel training pairs by generating synthetic speech with different speaker characteristics, creating (\textit{real source}, \textit{real reference}, \textit{synthetic target}) triplets~\cite{li2025starvc, tu2025o_o}. 
However, since the model learns to reproduce these synthetic targets, its output quality is inherently bounded by the external system's capabilities.
Recent advances attempt to preserve real speech as targets to overcome this ceiling.
SeedVC~\cite{liu2024zero} employs an external voice conversion system~\cite{qin2023openvoice} to generate timbre-perturbed inputs paired with real \textit{targets}, but initially focuses solely on timbre transfer rather than complete speaking style imitation, and requires an external model during training.
SynthVC~\cite{guo2025synthvc} trains on synthetic sources with real \textit{targets}, yet is limited to a fixed set of speakers and exhibits lower naturalness than its external VC system~\cite{liu2024zero}.
These limitations motivate the need for a zero-shot voice imitation approach that learns from real speech distributions without sacrificing naturalness.

To address these limitations, we introduce a novel data construction strategy: inverting the role of synthetic speech from training targets to training sources.
Specifically, given a real utterance spoken by speaker A, we use a TTS model~\cite{du2024cosyvoice} to re-synthesize the same linguistic content in a different voice (speaker B).
We then construct training triplets where the synthetic utterance serves as the \textit{source}, another real recording from speaker A serves as the \textit{reference}, and the original real utterance becomes the \textit{target}.
The model learns to transform the synthetic source to match the reference speaker's voice while preserving content.
This construction is valid because the source and target share identical textual content by design.
By learning to generate real human speech rather than synthetic outputs, MimicLM breaks the quality ceiling that constrains methods trained on synthetic \textit{targets}.
Moreover, since both \textit{reference} and \textit{target} come from real recordings of the same speaker, the model implicitly learns to capture and transfer voice characteristics without requiring explicit feature extraction or disentanglement.
However, this inversion introduces new challenges: training on synthetic inputs while performing inference on real inputs creates a distributional gap that leads to systematic performance degradation in real-world conditions.
Additionally, voice imitation itself is inherently more challenging than timbre-only voice conversion: transforming both timbre and prosodic patterns simultaneously alters the temporal structure of speech, making it harder to preserve linguistic content and often resulting in higher word error rates (WER). This elevation is an inherent characteristic of voice imitation, also observed in prior work~\cite{zhang2025vevo}.

To address these challenges, we propose MimicLM, an end-to-end voice imitation model incorporating two key techniques.
First, to mitigate content corruption when transforming complete speaking styles, we incorporate interleaved text-audio modeling: text tokens are interleaved with audio tokens in the input sequence, providing explicit content anchors that guide the model to preserve semantic information during voice transformation.
Second, to bridge the synthetic-to-real gap~\cite{su2024task}, we apply preference alignment during post-training.
Specifically, we conduct post-training by feeding the model real \textit{source-reference} pairs, sampling multiple candidate outputs, and ranking them by WER.
These ranked candidates form preference pairs that guide the model to generate content-faithful, natural outputs when processing real speech, effectively adapting it from the synthetic-input training regime to real-world conditions.

Experimental results demonstrate that MimicLM achieves strong naturalness while maintaining competitive similarity to state-of-the-art voice imitation systems. 
Notably, preference alignment yields a significant reduction in WER on real inputs, effectively bridging the distributional gap. 
Furthermore, scaling analysis reveals consistent improvements as data volume increases, suggesting substantial potential for further gains with larger datasets.
Our main contributions are: 
(1) We propose a role-swapping data construction strategy that uses TTS-generated speech as \textit{source} inputs and real recordings as \textit{targets}, enabling scalable training while learning from high-quality real speech distributions. 
(2) We introduce interleaved text-audio modeling and preference alignment to address the intelligibility degradation and distributional gap inherent in role-swapped training. 
(3) Through extensive evaluation, we demonstrate that this unified framework achieves competitive performance across naturalness, intelligibility, and similarity, offering a conceptually simpler alternative to complex disentanglement-based architectures.

\begin{figure*}[t]
  \includegraphics[width=0.9\textwidth]{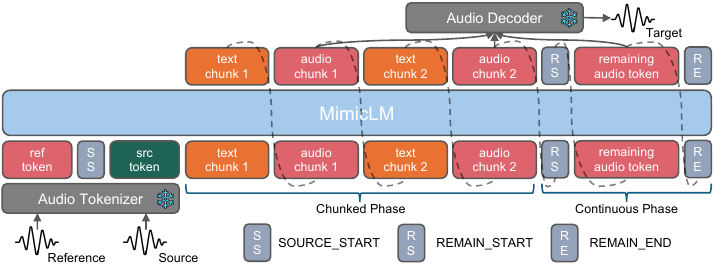}
  \centering
  \caption{Overview of MimicLM architecture for voice imitation. The model processes \textit{reference} audio (\textit{ref token}) for target timbre/style and \textit{source} audio (\textit{src token}), then generates the conversion \textit{target} in two phases: a chunked phase with interleaved text-audio prediction, followed by a continuous phase for remaining speech tokens. Both the audio tokenizer and decoder are frozen during training (snowflake icons). Special control tokens are annotated at bottom right.}
  \label{fig:model_structure}
\end{figure*}

\section{Related Work}
\label{sec:related_work}

\subsection{From Timbre to Voice Imitation}
\label{subsec:para1}
Voice conversion (VC) traditionally transforms only speaker timbre while 
preserving content and prosody~\cite{sisman2020overview}, whereas voice 
imitation (VI) reproduces both timbre and speaking style~\cite{zhang2025vevo}.

Early VI work used sequence-to-sequence models~\cite{zhang2019non, wang2023lm} 
or explicit prosody modeling~\cite{choi2023diff, lee2025hierspeech++}. 
Recently, zero-shot TTS~\cite{wang2023neural, anastassiou2024seed, wang2024maskgct, ju2024naturalspeech, du2024cosyvoice} has become dominant but requires external ASR~\cite{radford2022whisper} for speech-to-speech scenarios, introducing latency and error propagation.

Recent work has revisited VI from the VC perspective through direct speech-to-speech transformation. Vevo~\cite{zhang2025vevo} achieves state-of-the-art results using VQ-VAE information bottlenecks~\cite{van2017neural}. Similarly, SeedVC~\cite{liu2024zero} introduced a new version with explicit quantization and ASR-supervised feature learning to better support VI tasks. We explore whether simpler end-to-end VC architectures can achieve voice imitation by rethinking training data construction and model design.

\subsection{Addressing Parallel Data Scarcity}
\label{subsec:para2}
A fundamental bottleneck in training speech-to-speech conversion models is the scarcity of naturally parallel data. Recent approaches construct pseudo-parallel training pairs through two strategies.

\paragraph{Learning from Synthetic Targets.} 
StarVC~\cite{li2025starvc} employs a VC system~\cite{qin2023openvoice} to generate speaker variations, while O\_O-VC~\cite{tu2025o_o} uses multi-speaker TTS~\cite{kim2021conditional} to synthesize paired speech from the same linguistic content. 
While enabling scalable training, these approaches inherit quality limitations from their external synthesis systems.

\paragraph{Learning from Real Targets.} 
Recent work preserves real speech as training \textit{targets}. 
SeedVC~\cite{liu2024zero} applies timbre perturbation via external VC~\cite{qin2023openvoice}, reducing timbre leakage but requiring auxiliary models. 
SynthVC~\cite{guo2025synthvc} trains on synthetic sources with real \textit{targets} but supports only fixed speaker sets rather than zero-shot conversion.
We instead propose zero-shot voice imitation that learns from real \textit{targets} while capturing both timbre and speaking style, achieving natural output unconstrained by external TTS quality.

\begin{figure*}[t]
  \centering
  \includegraphics[width=0.90\textwidth]{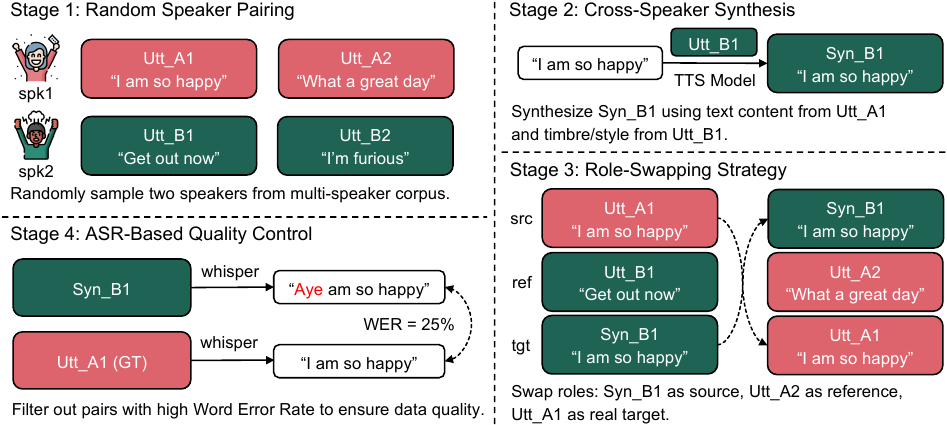}
  \caption{Four-stage pipeline for pseudo-parallel data construction. We randomly sample two speakers with their utterances (Stage 1), synthesize cross-speaker audio using TTS (Stage 2), apply role-swapping to ensure real speech serves as training \textit{target} (Stage 3), and filter low-quality pairs via ASR-based verification (Stage 4).}
  \label{fig:four_stages}
\end{figure*}

\subsection{Disentanglement or End-to-End Learning}
\label{subsec:para3}
Current approaches rely on explicit disentanglement through knowledge distillation~\cite{polyak2021speech, ju2024naturalspeech}, information bottlenecks~\cite{qian2019autovc, zhang2025vevo}, or acoustic perturbation~\cite{choi2021neural}. 
However, complete disentanglement remains difficult and requires multi-stage training with adversarial objectives~\cite{ju2024naturalspeech, lajszczak2024base}.

The availability of pseudo-parallel data enables end-to-end learning that bypasses explicit disentanglement. 
Large-scale models~\cite{wang2023neural, le2023voicebox, anastassiou2024seed} demonstrate that in-context learning can implicitly capture speaker characteristics when sufficient paired data is available.

\section{MimicLM}
\label{sec:method}

    \subsection{Overview}
    As illustrated in Figure~\ref{fig:model_structure}, our system consists of three components: (1) a frozen audio tokenizer that converts waveforms to discrete tokens, (2) a decoder-only Transformer that transforms source tokens to target tokens conditioned on reference, and (3) a flow-matching-based decoder that reconstructs waveforms. We adopt the tokenizer and decoder from CosyVoice 2.0~\cite{du2024cosyvoice}, which produces 25 tokens per second.
    Our approach achieves effective voice imitation through three key designs: (1) a role-swapping strategy that constructs pseudo-parallel training pairs with real speech as targets (Section~\ref{method:role-swapping}), (2) interleaved text-audio modeling that enhances content fidelity through explicit textual representation (Section~\ref{method:interleaved}), and (3) preference alignment that bridges the synthetic-to-real distribution gap during inference (Section~\ref{method:dpo}). 
    
    \subsection{Pseudo-Parallel Data Construction}
    \label{method:role-swapping}
    \paragraph{Motivation.}
    Parallel voice conversion corpora, recordings of the same content spoken by different speakers, are scarce~\cite{yoshino2016parallel}, and manual collection is expensive and time-consuming. Recent zero-shot TTS models~\cite{wang2024maskgct, du2024cosyvoice, xie2025fireredtts, zhou2025indextts2} offer a scalable alternative through synthetic data generation, but using synthetic speech as training objects introduces a quality ceiling. We address this through a role-swapping strategy that uses real speech as targets while leveraging TTS for content transformation.
        
    \paragraph{Four-Stage Pipeline.} 
    As shown in Figure~\ref{fig:four_stages}, our data construction pipeline comprises four stages:
    
    \textbf{Stage 1: Random Speaker Pairing.}
    From the filtered Emilia dataset~\cite{he2024emilia} containing over 620K English-speaking speakers (each with at least four utterances), we randomly sample two speakers, denoted as \texttt{spk1} and \texttt{spk2}, and retrieve three utterances: two adjacent utterances \texttt{Utt\_A1} and \texttt{Utt\_A2} from \texttt{spk1}, and one utterance \texttt{Utt\_B1} from \texttt{spk2}.
    
    \textbf{Stage 2: Cross-Speaker Synthesis.}
    We employ CosyVoice 2.0~\cite{du2024cosyvoice}, a zero-shot TTS model, to synthesize speech \texttt{Syn\_B1} using the text content of \texttt{Utt\_A1} (e.g., ``I am so happy'') and the timbre/style reference from \texttt{Utt\_B1}. A conventional approach would construct a training triplet (\texttt{Utt\_A1}, \texttt{Utt\_B1}, \texttt{Syn\_B1}), where the model learns to map \texttt{Utt\_A1} (\textit{source}) to \texttt{Syn\_B1} (\textit{target}) conditioned on \texttt{Utt\_B1} (\textit{reference}).
    
    However, this configuration suffers from two fundamental limitations: (1) \textit{Synthetic quality ceiling}: Training the model to generate synthetic speech creates an upper bound on output quality, as the targets themselves are imperfect despite advances in modern TTS systems. (2) \textit{Reference-target mismatch}: \texttt{Syn\_B1} may not accurately inherit the timbre and style from \texttt{Utt\_B1} due to inherent voice cloning errors in the TTS model, creating inconsistency between the target and reference during training.

    \textbf{Stage 3: Role-Swapping Strategy.}
    To address these issues, we propose a role-swapping strategy that inverts the conventional data configuration. The key insight is: instead of teaching the model to convert real speech \texttt{Utt\_A1} into synthetic speech \texttt{Syn\_B1}, we teach it to convert synthetic speech \texttt{Syn\_B1} back into real speech \texttt{Utt\_A1}.
    
    Concretely, we construct the training triplet as (\texttt{Syn\_B1}, \texttt{Utt\_A2}, \texttt{Utt\_A1}), where \texttt{Syn\_B1} serves as the \textit{source} (synthetic speech providing content), \texttt{Utt\_A2} as the \textit{reference} (real speech from \texttt{spk1} providing target timbre/style), and \texttt{Utt\_A1} as the \textit{target} (real speech from \texttt{spk1} as ground truth output).

    This inversion is valid because \texttt{Syn\_B1} and \texttt{Utt\_A1} share the same textual content by construction (Stage 2), making the task equivalent to voice conversion. This design brings two key advantages: (1) \textit{Real speech targets}: The model learns to generate real human speech directly, removing the quality ceiling imposed by synthetic training objects and potentially exceeding the quality of the TTS system used for data construction. (2) \textit{Better reference alignment}: Since \texttt{Utt\_A1} (\textit{target}) and \texttt{Utt\_A2} (\textit{reference}) are both from \texttt{spk1}, they naturally share the same speaker identity and exhibit similar timbre and style characteristics. This reduces the reference-target mismatch present in the conventional approach, where the synthetic \textit{target} may deviate from the \textit{reference} due to TTS cloning errors.
    
    \textbf{Stage 4: ASR-Based Quality Control.}
    Prior to ASR filtering, we apply voice activity detection (VAD)\footnote{\url{https://huggingface.co/nvidia/frame_vad_multilingual_marblenet_v2.0}} to trim leading and trailing non-speech segments from the synthesized utterances. This step addresses a common artifact in TTS-generated speech, where variable-length silence may precede the actual spoken content.
    We then apply Whisper-large-v3~\cite{radford2022whisper} to transcribe both \texttt{Syn\_B1} and \texttt{Utt\_A1}, retaining only pairs with WER below 0.1. This filtering removes 33\% of the data, yielding 8.5M high-quality training triplets (approximately 18K hours). We chose this threshold to balance data quantity and quality; a stricter threshold of 0.01 would remove 61\% of the data.

\subsection{Interleaved Text-Audio Modeling}
\label{method:interleaved}
\paragraph{Motivation.}
Recent work in speech language models has demonstrated that incorporating text prediction as an auxiliary task significantly improves speech intelligibility~\cite{xie2024mini, ding2025kimi, coreteam2025mimoaudio}. For voice imitation tasks, this is particularly important: unlike timbre-only conversion, imitating speaking style while preserving content fidelity presents greater challenges to maintaining intelligibility. We therefore adopt an interleaved text-audio architecture where textual predictions provide semantic guidance during audio synthesis. As shown in Table~\ref{tab:ablation}, this design substantially reduces WER compared to audio-only training.

\paragraph{Interleaved Sequence Construction.}
We extend Qwen2's~\cite{yang2024qwen2technicalreport} vocabulary with 6,561 speech tokens from CosyVoice 2.0's frozen audio tokenizer~\cite{du2024cosyvoice} and special control tokens for sequence management. As illustrated in Figure~\ref{fig:model_structure}, the input sequence consists of three components: (1) \textit{reference} tokens providing target timbre/style, (2) \textit{source} tokens prefixed by \texttt{<|SOURCE\_START|>}, and (3) interleaved text-audio chunks representing the conversion \textit{target}. This sequence operates in two phases:

\textbf{Chunked Phase.} We alternate between text chunks and audio chunks with sizes $C_{\text{text}}=5$ and $C_{\text{audio}}=25$ respectively. Each text chunk is enclosed by \texttt{<|TEXT\_START|>} and \texttt{<|TEXT\_END|>} tokens (omitted in Figure~\ref{fig:model_structure} for clarity), immediately followed by the corresponding audio chunk. This 1:5 ratio is deliberately designed: while natural temporal correspondence exhibits approximately three text tokens per 25 audio tokens (reflecting semantic density differences), we increase text chunk size to five, ensuring text predictions temporally lead audio synthesis. This temporal offset allows the model to leverage richer textual context as guidance for more intelligible audio generation.

\textbf{Continuous Phase.} After the chunked phase, any remaining content is generated continuously: remaining text tokens between \texttt{<|REMAIN\_START|>} and \texttt{<|TEXT\_END|>}, followed by remaining audio tokens ending with \texttt{<|REMAIN\_END|>}. This two-phase design accommodates variable-length inputs while maintaining structured guidance during the critical chunked generation phase.

\paragraph{Dual-Task Learning.}
The model is trained to simultaneously predict text and audio tokens. For loss computation, control tokens following text chunks (\texttt{<|TEXT\_END|>}) contribute to text loss, while those following audio tokens (\texttt{<|TEXT\_START|>}, \texttt{<|REMAIN\_START|>}, \texttt{<|REMAIN\_END|>}) contribute to audio loss. 

The training objective is:
\begin{equation}
\mathcal{L} = 0.5 \mathcal{L}_{\text{text}} + 0.5 \mathcal{L}_{\text{audio}},
\end{equation}
where both losses are cross-entropy computed at their respective token positions.

\subsection{Preference Alignment}
\label{method:dpo}
\paragraph{Motivation.}
While role-swapping ensures high-quality \textit{target} speech via real recordings, the \textit{source} speech remains synthetic during training. This creates a distributional mismatch at inference time. As shown in Table~\ref{tab:sim2real}, our SFT model achieves 4.30\% WER when evaluated on Syn/Real pairs (matching the training distribution), but performance degrades to 15.80\% WER on Real/Real pairs (real-world inference scenario). To bridge this synthetic-to-real gap~\cite{su2024task}, we employ Direct Preference Optimization (DPO)~\cite{rafailov2023direct} to align the model with real-world acoustic conditions.

\paragraph{Preference Data Construction.}
We construct two-stage preference data from 150K speaker pairs in Emilia. For each \textit{(source, reference)} pair, we generate $K=8$ candidate outputs using nucleus sampling and rank them using automatic metrics (Section~\ref{exp:metrics}). Stage 1 uses the base model to generate candidates and prioritizes speech intelligibility (WER) to align with real source characteristics. After Stage 1 optimization, Stage 2 generates new candidates and focuses on acoustic similarity including voice timbre, accent, and emotional expression. Full construction details are in Appendix~\ref{app:dpo_details}.

\paragraph{Training Objective.}
Following~\cite{rafailov2023direct}, the DPO loss is:
\begin{equation}
\footnotesize
\mathcal{L}_{\text{DPO}} = -\mathbb{E}_D \Big[ \log \sigma \Big( \beta \big( \log \tfrac{\pi_\theta(y_w|x)}{\pi_{\text{ref}}(y_w|x)} - \log \tfrac{\pi_\theta(y_l|x)}{\pi_{\text{ref}}(y_l|x)} \big) \Big) \Big],
\end{equation}
where $\pi_\theta$ is the policy being optimized, $\pi_{\text{ref}}$ is the frozen reference model (our SFT model), and $(x, y_w, y_l)$ denotes the (input, chosen, rejected) triplet. We set $\beta=0.1$ following the default configuration in~\cite{rafailov2023direct}.

\begin{table*}[t]
\centering
\caption{Performance comparison on SeedTTS \textit{test-vc-en}. We compare our method against two groups of baselines: (1) \textit{Timbre-only} systems, and (2) \textit{Full Voice Imitation} systems (our direct baselines: Vevo and SeedVC v2), which transfer both timbre and style. ``$\uparrow$'' indicates higher is better, and ``$\downarrow$'' indicates lower is better. WER comparison is only made among voice imitation systems.}
\label{tab:main_results}
\renewcommand{\arraystretch}{1.0}
\resizebox{\textwidth}{!}{%
    \begin{tabular}{lcccccccc}
        \toprule
        \multirow{2}{*}{\textbf{Model}} & \multicolumn{4}{c}{\textbf{Naturalness \& Speech Quality}} & \multicolumn{1}{c}{\textbf{Intelligibility}} & \multicolumn{3}{c}{\textbf{Speaker Similarity}} \\
        \cmidrule(lr){2-5} \cmidrule(lr){6-6} \cmidrule(lr){7-9}
         & UTMOS $\uparrow$ & OVRL $\uparrow$ & SIG $\uparrow$ & BAK $\uparrow$ & WER (\%) $\downarrow$ & S-SIM $\uparrow$ & A-SIM $\uparrow$ & E-SIM $\uparrow$ \\
        \midrule
        \rowcolor{gray!10}
        \multicolumn{9}{l}{\textit{\textbf{Timbre-only Voice Conversion}}} \\
        FACodec~\cite{ju2024naturalspeech} & 2.13 & 3.65 & 4.22 & 3.87 & 4.86 & 0.372 & 0.571 & 0.912 \\
        MeanVC~\cite{ma2025meanvc} & 2.84 & 3.65 & 4.04 & 4.14 & 17.82 & 0.419 & 0.450 & 0.909 \\
        OpenVoice v2~\cite{qin2023openvoice} & 2.55 & \textbf{4.15} & \textbf{4.47} & \underline{4.43} & 4.15 & 0.424 & 0.552 & 0.918 \\
        LSCodec~\cite{guo2024lscodec} & 2.84 & 3.94 & 4.33 & 4.25 & 8.76 & 0.445 & 0.583 & 0.911 \\
        CosyVoice 2.0~\cite{du2024cosyvoice} & 3.04 & 3.98 & 4.31 & 4.38 & 4.28 & 0.539 & 0.647 & 0.919 \\
        SeedVC~\cite{liu2024zero} & 2.79 & 3.71 & 4.19 & 4.03 & 3.25 & 0.587 & 0.684 & 0.922 \\
        \midrule
        \rowcolor{gray!10}
        \multicolumn{9}{l}{\textit{\textbf{Full Voice Imitation (Timbre + Style Transfer)}}} \\
        SeedVC v2~\cite{liu2024zero} & 2.94 & 3.65 & 4.14 & 4.01 & \textbf{6.32} & 0.553 & 0.653 & 0.917 \\
        Vevo~\cite{zhang2025vevo} & 2.83 & 3.77 & 4.27 & 4.00 & 9.10 & \textbf{0.652} & \textbf{0.727} & \textbf{0.926} \\
        Ours (SFT) & \textbf{3.31} & \underline{4.12} & 4.43 & 4.42 & 12.80 & 0.571 & 0.692 & 0.912 \\
        Ours (DPO) & \underline{3.22} & \textbf{4.15} & \underline{4.45} & \textbf{4.45} & \underline{8.25} & \underline{0.601} & \underline{0.699} & \underline{0.925} \\
        \bottomrule
    \end{tabular}%
}
\end{table*}

\begin{table}[h]
    \centering
    \caption{Subjective evaluation results with Mean Opinion Scores (MOS) on a 1--5 scale. Scores are reported as mean $\pm$ 95\% confidence interval.}
    \label{tab:mos}
    \resizebox{\columnwidth}{!}{%
        \begin{tabular}{l|cccc}
        \toprule
        \textbf{Model} & \textbf{N-MOS} & \textbf{S-MOS} & \textbf{A-MOS} & \textbf{E-MOS} \\
        \midrule
        SeedVC v2  & 3.14 $\pm$ 0.11 & 3.03 $\pm$ 0.12 & 3.82 $\pm$ 0.12 & 3.61 $\pm$ 0.16 \\
        Vevo       & 3.85 $\pm$ 0.14 & 4.32 $\pm$ 0.13 & \textbf{4.64 $\pm$ 0.09} & \textbf{4.23 $\pm$ 0.09} \\
        Ours (DPO) & \textbf{4.71 $\pm$ 0.08} & \textbf{4.62 $\pm$ 0.10} & 4.53 $\pm$ 0.11 & 3.94 $\pm$ 0.13 \\
        \bottomrule
        \end{tabular}%
    }
\end{table}

\begin{table}[h]
    \centering
    \small
    \setlength{\tabcolsep}{4pt}
    \caption{Synthetic-to-real gap analysis on MimicLM-Test. Column headers indicate \textit{source/reference} input types, where \textit{source} provides linguistic content and \textit{reference} provides target voice. All values are WER (\%).}
    \label{tab:sim2real}
    \begin{tabular}{l|ccc}
    \toprule
    \textbf{Model} & \textbf{Real/Real} & \textbf{Syn/Real} & \textbf{Real/Syn} \\
    \midrule
    Vevo & 17.99 & 13.90 & 20.44 \\
    Ours (SFT) & 15.80 & 4.30 & 18.48 \\
    Ours (DPO) & \textbf{13.81} & \textbf{3.63} & \textbf{15.58} \\
    \bottomrule
    \end{tabular}
\end{table}

\section{Experiments and Results}
\label{sec:experiments}

\subsection{Experimental Settings}

    \paragraph{Training Data.}
    We construct a pseudo-parallel corpus following Section~\ref{method:role-swapping} based on Emilia~\cite{he2024emilia}, a large-scale multilingual speech dataset with 1.13M English and 0.91M Chinese speakers. We select 620K speakers per language (each with at least four utterances), generating 8.5M English pairs (18K hours) and 0.74M Chinese pairs (1.6K hours). All experiments use English data unless noted; multilingual training results are in Appendix~\ref{app:multilingual}.
    
    \paragraph{Evaluation Data.}
    We evaluate on two benchmarks: (1) SeedTTS \textit{test-vc-en}~\cite{anastassiou2024seed}: A zero-shot TTS/VC benchmark from Common Voice~\cite{ardila2020common}. We select all utterances longer than 4 seconds. (2) MimicLM-Test: To analyze the synthetic-to-real gap, we construct a diagnostic set with 833 speaker pairs (1,666 utterances) from Emilia. Each pair is evaluated under three conditions based on input types: \textit{Real/Real}, \textit{Syn/Real}, and \textit{Real/Syn}, where the format indicates \textit{Source/Reference} audio types.

    \paragraph{Evaluation Metrics.}
    \label{exp:metrics}
    We employ both objective and subjective metrics. For naturalness, we use DNSMOS\footnote{\url{https://github.com/microsoft/DNS-Challenge}}~\cite{reddy2021dnsmos} to evaluate speech quality (SIG), background noise (BAK), and overall quality (OVRL), and UTMOSv2\footnote{\url{https://github.com/sarulab-speech/UTMOSv2}}~\cite{baba2024t05} for automatic Mean Opinion Score (MOS) prediction. For intelligibility, we transcribe outputs with Whisper-Large-v3~\cite{radford2023robust} and compute Word Error Rate (WER). For similarity, we compute cosine similarity between embeddings of generated and reference audio using WavLM-Large\footnote{\url{https://github.com/BytedanceSpeech/seed-tts-eval}}~\cite{chen2022wavlm} for speaker similarity (S-SIM), CommonAccent~\cite{ardila2020common} for accent (A-SIM), and emotion2vec\footnote{\url{https://github.com/ddlBoJack/emotion2vec}}~\cite{ma2023emotion2vec} for emotion (E-SIM). We also conduct subjective evaluation where human raters score naturalness (N-MOS), speaker similarity (S-MOS), accent similarity (A-MOS), and emotion similarity (E-MOS) on a 1--5 scale. Details are in Appendix~\ref{app:subjective}.
    
    \paragraph{Baselines}
    We compare against open-source state-of-the-art zero-shot voice conversion systems in two categories. Timbre-only systems including CosyVoice 2.0~\cite{du2024cosyvoice}, FACodec~\cite{ju2024naturalspeech}, OpenVoice v2~\cite{qin2023openvoice}, LSCodec~\cite{guo2024lscodec}, and SeedVC~\cite{liu2024zero} transfer speaker identity while preserving source prosody. Full voice imitation systems including SeedVC v2~\cite{liu2024zero} and Vevo~\cite{zhang2025vevo} transfer both timbre and style.
    
    \paragraph{Training and Inference Setup}
    We train the model in two stages on NVIDIA A800 GPUs. Stage 1 performs supervised fine-tuning for 4 epochs with effective batch size of 128, learning rate $5 \times 10^{-4}$, and warmup ratio 0.03. Stage 2 applies DPO alignment with effective batch size of 32, learning rate $1 \times 10^{-5}$, $\beta=0.1$, and warmup ratio 0.05 for 4 epochs.
    We perform inference in bfloat16 precision. Text generation uses temperature 0.7, top-p 0.92, and repetition penalty 1.05, while audio generation uses temperature 0.8, top-p 0.9, and repetition penalty 1.2. More details are in Appendix~\ref{app:training_details}.

\begin{table*}[tp]
  \centering
  \small
  \renewcommand{\arraystretch}{1.0}
  \caption{Ablation study on key components using \textit{base}-scale data. RS = Role-Swapping, IT = Interleaved Text modeling. The baseline uses standard triplets (\texttt{Utt\_A1}, \texttt{Utt\_B1}, \texttt{Syn\_B1}) without IT. RS uses role-swapped triplets (\texttt{Syn\_B1}, \texttt{Utt\_A2}, \texttt{Utt\_A1}). All configurations are evaluated on SeedTTS \textit{test-vc-en}. Training hyperparameters follow the SFT stage setup described in Appendix~\ref{app:training_details}.}
  \label{tab:ablation}
  \begin{tabular}{l|ccc|c|ccc}
    \toprule
    \textbf{Configuration} & \textbf{OVRL $\uparrow$} & \textbf{SIG $\uparrow$} & \textbf{BAK $\uparrow$} & \textbf{WER (\%) $\downarrow$} & \textbf{S-SIM $\uparrow$} & \textbf{A-SIM $\uparrow$} & \textbf{E-SIM $\uparrow$} \\
    \midrule
    w/o RS, w/o IT & 3.99 & 4.39 & 4.25 & 18.25 & 0.547 & 0.678 & 0.903 \\
    w/ RS, w/o IT & 4.05 & 4.41 & 4.33 & 20.69 & 0.555 & 0.684 & \underline{0.910} \\
    w/o RS, w/ IT & 4.03 & 4.41 & 4.31 & \underline{15.34} & 0.547 & 0.681 & 0.896 \\
    w/ RS, w/ IT (SFT) & \underline{4.11} & \underline{4.43} & \underline{4.41} & 18.64 & \underline{0.560} & \textbf{0.691} & \textbf{0.913} \\
    SFT + DPO & \textbf{4.12} & \textbf{4.44} & \textbf{4.42} & \textbf{14.73} & \textbf{0.573} & \underline{0.688} & 0.905 \\
    \bottomrule
  \end{tabular}
\end{table*}

\subsection{Main Results}
\label{sec:main_results}

\paragraph{Comparison with Baselines.}
\label{sec:vs_baselines}
Table~\ref{tab:main_results} presents our method's performance on SeedTTS \textit{test-vc-en} against state-of-the-art baselines. Our SFT model is trained on 8.5M pseudo-parallel pairs (18K hours) constructed from real speech via role-swapping, while our DPO model further applies preference alignment as post-training. Overall, our method achieves competitive performance across naturalness, intelligibility, and similarity metrics.
(1) \textit{Naturalness and Speech Quality.} Our method demonstrates strong naturalness, with both SFT and DPO versions achieving competitive scores across quality metrics. After DPO alignment, these naturalness metrics remain stable, indicating that preference optimization maintains speech quality while improving other performance dimensions.
(2) \textit{Intelligibility.} Voice imitation systems generally exhibit higher WER than timbre-only approaches, as transforming both timbre and speaking style increases content preservation complexity. Among full imitation systems, our SFT model's WER falls between SeedVC v2 and Vevo. The DPO process substantially reduces WER, demonstrating that post-training alignment on real speech inputs effectively bridges the synthetic-to-real gap encountered during inference.
(3) \textit{Similarity Performance.} Our method achieves competitive similarity performance across all dimensions. For speaker identity similarity (S-SIM), our DPO model outperforms timbre-only VC systems and achieves results comparable to other full imitation methods, surpassing SeedVC v2 and approaching Vevo's performance. For accent similarity (A-SIM), our DPO model demonstrates advantages over timbre-only baselines, with performance positioned between SeedVC v2 and Vevo among full imitation systems. For emotion similarity (E-SIM), our DPO model closely matches the best-performing Vevo on this test set. The consistent improvements from SFT to DPO across all similarity metrics demonstrate that preference alignment effectively enhances the model's ability to capture comprehensive voice characteristics.
(4) \textit{Subjective Evaluation.}
We conduct human evaluation to complement the objective metrics. As shown in Table~\ref{tab:mos}, our DPO model achieves the highest scores in N-MOS and S-MOS. These subjective results validate the effectiveness of our approach.

\paragraph{Addressing Synthetic-to-Real Gap.}
\label{exp:syn2real}

Table~\ref{tab:sim2real} quantifies the distributional shift between training and inference conditions on our diagnostic set MimicLM-Test. As discussed in Section~\ref{method:dpo}, our SFT model performs well on Syn/Real pairs (matching the training distribution) but exhibits substantially higher WER on Real/Real pairs (real-world inference scenario), confirming the synthetic-to-real gap. DPO effectively bridges this gap, achieving the lowest WER across all input configurations and surpassing Vevo under Real/Real conditions despite Vevo being trained on real data. Notably, DPO maintains strong performance on Syn/Real pairs while successfully adapting to Real/Real pairs, demonstrating that preference optimization generalizes across different input distributions without sacrificing quality on the training domain.

\paragraph{Data Scaling.}
\begin{figure}[t]
    \centering
    \includegraphics[width=\columnwidth]{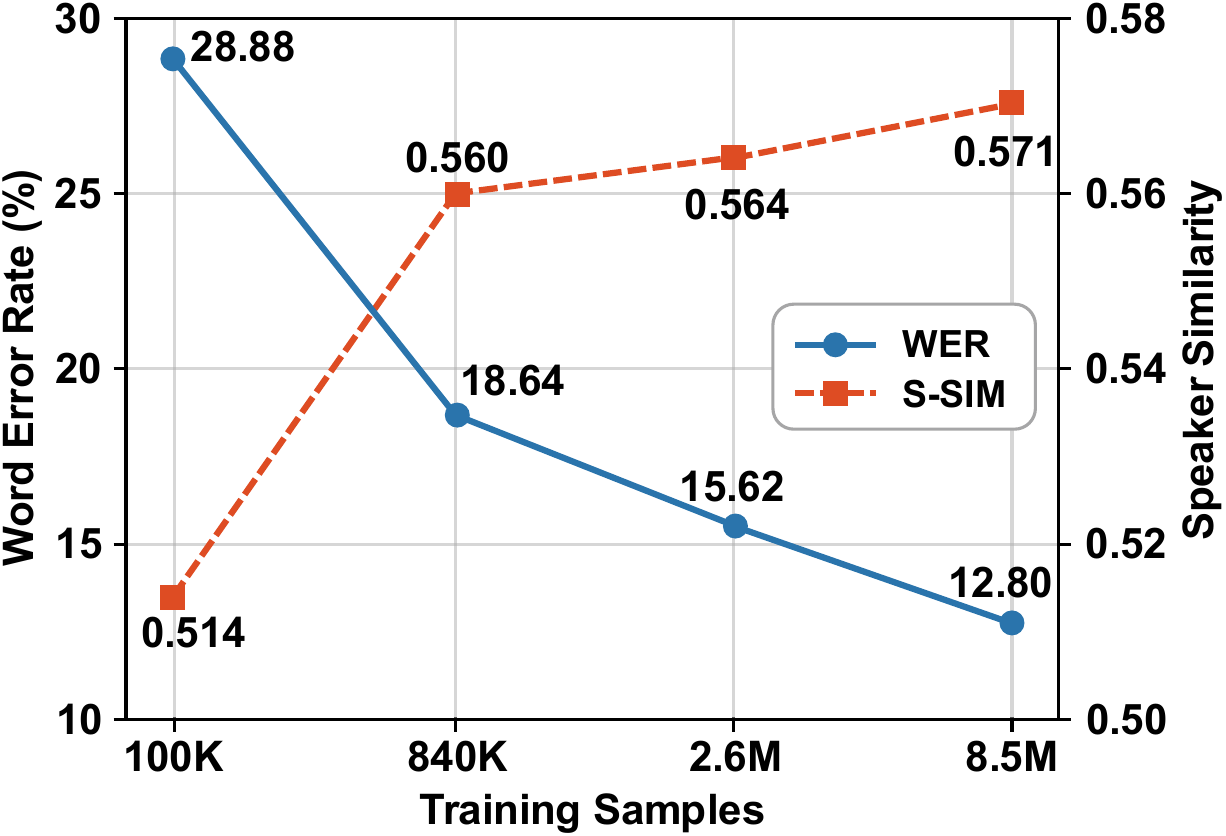}
    \caption{Impact of training data scale on WER and speaker similarity (S-SIM) evaluated on SeedTTS \textit{test-vc-en}.}
    \label{fig:scaling}
\end{figure}

To investigate scaling behavior, we train models on datasets of varying sizes: \textit{tiny} (100K samples), \textit{base} (840K), \textit{medium} (2.6M), and \textit{large} (8.5M), keeping architecture and hyperparameters fixed. Figure~\ref{fig:scaling} shows that performance consistently improves with scale. The transition from \textit{tiny} to \textit{base} demonstrates strong improvements in both WER and speaker similarity, indicating substantial benefits in the small-data regime. As data scale increases further, gains follow typical scaling law patterns with diminishing returns, though neither metric saturates at the largest scale, suggesting potential for further improvement with more data. Notably, WER exhibits steeper improvements than S-SIM across scales, suggesting that content preservation and voice characteristic modeling may have different scaling dynamics.

\subsection{Ablation Studies}
\label{sec:ablation}
We validate the effectiveness of role-swapping (RS), interleaved text modeling (IT), and DPO through systematic ablations on \textit{base}-scale data (840K samples). Table~\ref{tab:ablation} presents results on SeedTTS \textit{test-vc-en}. All ablated models are trained from scratch under identical training configurations, ensuring a fair comparison of each component's individual contribution.
(1) \textit{Role-swapping} (rows 1 vs. 2, rows 3 vs. 4) consistently improves naturalness and similarity, confirming that real target speech enhances voice characteristic transfer. 
(2) \textit{Interleaved text modeling} (rows 1 vs. 3, rows 2 vs. 4) substantially reduces WER by providing explicit linguistic anchors that prevent content collapse. 
Their combination in the SFT model achieves the best naturalness with strong similarity and reasonable WER, validating synergistic effects. 
(3) \textit{Preference alignment} further reduces WER and improves speaker similarity while maintaining naturalness. Note that this DPO uses \textit{base}-scale data optimized for WER and S-SIM, whereas our main model uses \textit{large}-scale data with multi-stage optimization (Section~\ref{method:dpo}).

\section{Conclusion}
We presented MimicLM, a voice imitation system that learns from real speech distributions by using synthetic speech as training sources rather than targets. This role-swapping data construction strategy breaks the quality ceiling imposed by synthesis systems while ensuring natural alignment between references and targets. Combined with interleaved text-audio modeling for content preservation and preference alignment for handling real speech inputs, MimicLM achieves competitive performance compared to state-of-the-art systems while offering a simpler alternative to complex disentanglement-based architectures. Our work demonstrates that inverting the role of synthetic data enables effective voice imitation without sacrificing naturalness or requiring elaborate architectural designs.

\section*{Limitations}
Our approach has several limitations worth noting. First, while avoiding synthetic \textit{targets}, our method still depends on TTS quality for generating training sources. Poor TTS outputs require filtering (33\% removal rate), potentially limiting data efficiency and introducing biases from the external synthesis system. Second, despite DPO alignment, our model exhibits higher WER than timbre-only systems when processing real inputs, indicating that the synthetic-to-real gap and the complexity of joint timbre-prosody transformation remain challenging. Third, our two-stage pipeline and large-scale data construction (8.5M pairs with TTS synthesis and preference optimization) demand substantial computational resources, which may limit accessibility for researchers with limited budgets. Finally, our evaluation focuses primarily on English with limited multilingual analysis. Real-world diversity in accents, speaking styles, and acoustic conditions may present challenges not fully covered by current benchmarks, warranting further investigation into edge cases and challenging scenarios.

\bibliography{custom}

\appendix

\section{Multilingual Training Results}
\label{app:multilingual}

To demonstrate language generalizability, we train models on Chinese data constructed following the same Role-Swapping procedure (Section~\ref{method:role-swapping}). We generate 740K Chinese pairs (1.6K hours) from Emilia and evaluate on SeedTTS \textit{test-zh}~\cite{anastassiou2024seed}, reporting Word Character Error Rate (CER) for Chinese. Following the English evaluation setup, we filter the test set to retain only examples with reference utterances longer than 4 seconds.

Table~\ref{tab:chinese_results} presents results for monolingual and multilingual training configurations. The Chinese-only model demonstrates our approach works effectively for this language. For multilingual training, outcomes vary with data balance: on Chinese evaluation, adding balanced English data improves performance, but disproportionately large English data causes degradation; on English evaluation, Chinese data slightly hurts performance with small English datasets but provides gains with larger ones.

\begin{table}[t]
\centering
\caption{Performance on Chinese (SeedTTS \textit{test-zh}) and English (SeedTTS \textit{test-vc-en}) test sets for different training data configurations.}
\label{tab:chinese_results}
\resizebox{\linewidth}{!}{%
\begin{tabular}{lcccc}
    \toprule
    \multirow{2}{*}{\textbf{Training Data}} & \multicolumn{2}{c}{\textbf{English}} & \multicolumn{2}{c}{\textbf{Chinese}} \\
    \cmidrule(lr){2-3} \cmidrule(lr){4-5}
     & WER (\%) $\downarrow$ & S-SIM $\uparrow$ & CER (\%) $\downarrow$ & S-SIM $\uparrow$ \\
    \midrule
    English-only (0.84M) & 18.64 & 0.560 & - & - \\
    English-only (2.6M) & 15.62 & 0.564 & - & - \\
    Chinese-only (0.74M) & - & - & 13.16 & 0.630 \\
    Chinese (0.74M) + English (0.84M) & 19.49 & 0.558 & 12.84 & 0.632 \\
    Chinese (0.74M) + English (2.6M) & 15.21 & 0.566 & 13.68 & 0.631 \\
    \bottomrule
\end{tabular}
}
\end{table}

\section{DPO Preference Pair Construction}
\label{app:dpo_details}
We conducted two rounds of DPO training, each utilizing 150,000 input samples from the training set. The preference pair construction process consists of two stages:

\textbf{Stage 1: Candidate Generation.} For each input sample, we employ a multi-process inference framework to generate 8 candidate responses (both text and audio). The generation process uses nucleus sampling with configurable temperature and top-p parameters for both text ($T_{\text{text}}=0.7$, $p_{\text{text}}=0.92$) and audio modalities ($T_{\text{audio}}=0.8$, $p_{\text{audio}}=0.9$). The framework supports distributed inference across 8 GPUs with checkpoint-based resumption capabilities.

\textbf{Stage 2: Pareto-Optimal Pair Selection.} We employ a multi-objective optimization strategy based on Pareto dominance to construct preference pairs. For each sample, all candidate pairs are evaluated using three metrics: (1) \textit{Audio WER} (lower is better), measuring speech recognition accuracy; (2) \textit{SIM} (higher is better), measuring overall similarity to the reference; and (3) \textit{eSIM} (higher is better), capturing fine-grained similarity. 

A candidate $c_1$ is considered to dominate $c_2$ (forming a chosen-rejected pair) if and only if:
\begin{itemize}[leftmargin=*,noitemsep,topsep=2pt]
    \item $c_1$ is no worse than $c_2$ on all metrics (Pareto condition);
    \item $c_1$ is strictly better than $c_2$ on at least one metric;
    \item The improvement on each metric exceeds a minimum threshold $\delta_{\text{min}}$;
    \item Both $c_1$ (chosen) and $c_2$ (rejected) satisfy quality constraints.
\end{itemize}

\noindent\textbf{Filtering Criteria.} We denote the minimum improvement threshold as $\delta_{\text{min}}$, the maximum allowed value for chosen samples as $v^c_{\text{max}}$, and the maximum allowed value for rejected samples as $v^r_{\text{max}}$. The specific criteria are:

\noindent\textit{Round 1:} Audio WER: $\delta_{\text{min}}=0.05$, $v^c_{\text{max}}=0.30$, $v^r_{\text{max}}=0.60$; SIM: $\delta_{\text{min}}=0.01$, no $v^c_{\text{max}}$ or $v^r_{\text{max}}$ constraints; eSIM: $\delta_{\text{min}}=0.01$, no $v^c_{\text{max}}$ or $v^r_{\text{max}}$ constraints.

\noindent\textit{Round 2:} Audio WER: $\delta_{\text{min}}=0.00$, $v^c_{\text{max}}=0.30$, $v^r_{\text{max}}=0.60$; SIM: $\delta_{\text{min}}=0.02$, no constraints; eSIM: $\delta_{\text{min}}=0.02$, no constraints.

Round 1 enforces stricter minimum improvements ($\delta_{\text{min}}=0.05$ for WER) to ensure high-quality pairs, while Round 2 relaxes the WER threshold to 0.00 but increases similarity requirements to 0.02, encouraging more diverse training signals. The quality constraints ensure that chosen samples maintain low WER ($\leq 0.30$) while preventing overly poor rejected samples (WER $\leq 0.60$).

\section{Subjective Evaluation Protocol}
\label{app:subjective}

We conduct subjective evaluation on a randomly selected subset of 20 audio pairs from the SeedTTS \textit{test-vc-en} dataset. Each sample is evaluated by 10 trained listeners using a 5-point Likert scale (1: Bad, 2: Poor, 3: Fair, 4: Good, 5: Excellent). 

The evaluation covers four dimensions: Naturalness (N-MOS), Speaker Similarity (S-MOS), Accent Similarity (A-MOS), and Emotion Similarity (E-MOS). For N-MOS, listeners assess the overall naturalness and audio quality of the generated speech, disregarding any differences in speaker characteristics or content accuracy. For S-MOS, listeners evaluate how closely the generated speech matches the target speaker's timbre and voice characteristics in the reference prompt, while disregarding differences in content or audio quality. For A-MOS, listeners assess how well the generated speech preserves the accent and pronunciation patterns of the target speaker. For E-MOS, listeners evaluate how closely the generated speech matches the emotional tone and expressiveness of the target speaker. For each dimension, listeners are instructed to focus solely on that specific aspect while ignoring other characteristics. The final MOS score for each model is computed as the average of all listener ratings, with confidence intervals calculated based on the standard error.

\section{Training Details}
\label{app:training_details}

\paragraph{Stage 1: Supervised Fine-Tuning}
We train the MimicLM model based on the Qwen2.5-0.5B~\cite{yang2024qwen2technicalreport} architecture on 8 NVIDIA A800 GPUs for 4 epochs. The training configuration is as follows: per-device batch size of 4 with gradient accumulation steps of 4, resulting in an effective batch size of 128. We employ the AdamW optimizer with a learning rate of $5 \times 10^{-4}$, $\beta_2=0.999$, weight decay of 0.01, warmup ratio of 0.03, and cosine learning rate scheduling. The model uses Flash Attention 2~\cite{dao2023flashattention2} and gradient checkpointing for training efficiency. The maximum sequence length is set to 2560 tokens, and the text and audio loss weights are both set to 0.5. Text chunks and audio chunks are configured with sizes of 5 and 25 tokens, respectively.

\paragraph{Stage 2: DPO Alignment}
We perform Direct Preference Optimization (DPO) training on 4 GPUs using the checkpoint from Stage 1. The training runs for 4 epochs with per-device batch size of 4 and gradient accumulation steps of 2, yielding an effective batch size of 32. We reduce the learning rate to $1 \times 10^{-5}$ with the DPO $\beta$ parameter set to 0.1, warmup ratio of 0.05, weight decay of 0.01, and maximum gradient norm clipping of 1.0. Both training stages utilize bfloat16 mixed precision training.

\section{Broader Impact and Potential Risks}
\label{app:risks}

While MimicLM advances voice imitation technology for legitimate applications such as personalized voice assistants, audiobook narration, and accessibility tools, we acknowledge potential misuse risks that warrant careful consideration.

The primary concern is unauthorized voice cloning for impersonation or fraud. Our model can generate speech that imitates a target speaker's voice using only a 3-second prompt, which could potentially be exploited for social engineering attacks, spreading misinformation through fake audio, or creating non-consensual voice replicas. Although our system is designed for research purposes and legitimate use cases, malicious actors could adapt similar techniques for harmful purposes.

To mitigate these risks, we recommend several safeguards for practical deployment. First, implement speaker verification and consent mechanisms before allowing voice cloning of any individual. Second, incorporate watermarking or fingerprinting techniques to trace generated audio back to its source. Third, develop and deploy detection systems capable of identifying synthetic speech generated by voice imitation models. Fourth, establish clear usage policies and legal frameworks that criminalize unauthorized voice cloning and impersonation.

We emphasize that voice imitation technology itself is neutral—its impact depends on how it is deployed and regulated. The research community, policymakers, and technology developers must work collaboratively to ensure these technologies are used responsibly. We release our work to advance scientific understanding while encouraging ongoing dialogue about ethical deployment practices and appropriate regulatory frameworks.

Our model's current limitations (higher WER on real inputs, dependency on high-quality prompts) provide some natural barriers against trivial misuse, but these technical limitations should not be relied upon as primary safeguards. As the technology matures, proactive measures become increasingly important to prevent harm while preserving beneficial applications.

\section{Use of Large Language Models}
During the preparation of this manuscript, a Large Language Model (LLM) was utilized as a writing
aid to improve the overall linguistic quality and clarity. This assistance was confined to copy-editing
tasks, such as correcting grammatical and spelling errors, rephrasing sentences for enhanced flow
and readability, and ensuring conciseness. All scientific contributions, including the research ideas,
experimental design, analysis, and conclusions presented herein, are entirely the original work of the
human authors.

\section{TTS Model Selection}
\label{app:tts_selection}

To select the TTS system used for cross-speaker synthesis in Stage 2, we evaluated several state-of-the-art open-source zero-shot TTS models on the SeedTTS \textit{test-en} and \textit{test-zh} benchmarks~\cite{anastassiou2024seed}, following the evaluation protocol in~\cite{wang2025tadicodec}. Table~\ref{tab:tts_selection} reports WER and speaker similarity (SIM) for each system.

\begin{table}[h]
\centering
\caption{Comparison of zero-shot TTS systems on SeedTTS \textit{test-en} and \textit{test-zh}. Results are from~\cite{wang2025tadicodec}. SIM denotes cosine speaker similarity.}
\label{tab:tts_selection}
\resizebox{\columnwidth}{!}{%
\begin{tabular}{lccccc}
\toprule
\multirow{2}{*}{\textbf{System}} & \multirow{2}{*}{\textbf{Frame Rate}} & \multicolumn{2}{c}{\textbf{English}} & \multicolumn{2}{c}{\textbf{Chinese}} \\
\cmidrule(lr){3-4} \cmidrule(lr){5-6}
 & & WER\,$\downarrow$ & SIM\,$\uparrow$ & WER\,$\downarrow$ & SIM\,$\uparrow$ \\
\midrule
FireRedTTS~\cite{guo2024fireredtts} & 25 & 8.53 & 0.46 & 1.27 & 0.65 \\
SparkTTS~\cite{wang2025spark} & 50 & 2.50 & 0.57 & 1.78 & 0.66 \\
Llasa~\cite{ye2025llasa} (16\,kHz) & 50 & 3.94 & 0.58 & 8.02 & 0.64 \\
F5-TTS~\cite{chen2025f5} & 93.75 & 3.02 & 0.63 & 3.87 & 0.71 \\
CosyVoice\,2~\cite{du2024cosyvoice} & 25 & 2.89 & 0.66 & 1.29 & 0.76 \\
\bottomrule
\end{tabular}%
}
\end{table}

CosyVoice\,2 achieves the best bilingual speaker similarity while maintaining competitive WER on both test sets, and operates at a low frame rate of 25 Hz---reducing synthesis cost relative to higher-rate alternatives. These properties make it the most suitable choice for large-scale pseudo-parallel data construction in our pipeline.

\section{Audio Tokenizer Selection}
\label{app:tokenizer_selection}

We selected the audio tokenizer by evaluating multiple codec systems on the SeedTTS \textit{test-en} benchmark using a reconstruction protocol, following~\cite{wang2025tadicodec}. Table~\ref{tab:tokenizer_selection} reports WER, speaker similarity (SIM), and UTMOS for each tokenizer.

\begin{table}[h]
\centering
\caption{Comparison of audio tokenizers evaluated via speech reconstruction on SeedTTS \textit{test-en}. Results are from~\cite{wang2025tadicodec}. ``Semantic Distill'' indicates whether the tokenizer incorporates semantic distillation.}
\label{tab:tokenizer_selection}
\resizebox{\columnwidth}{!}{%
\begin{tabular}{lcccccc}
\toprule
\textbf{System} & \textbf{Rate (Hz)} & \textbf{Codebooks} & \textbf{Semantic} & \textbf{WER\,$\downarrow$} & \textbf{SIM\,$\uparrow$} & \textbf{UTMOS\,$\uparrow$} \\
\midrule
EnCodec~\cite{defossez2022high} & 75 & 2 & \texttimes & 5.36 & 0.48 & 1.54 \\
Mimi~\cite{defossez2024moshi} & 12.5 & 6 & \checkmark & 4.51 & 0.52 & 3.09 \\
BigCodec~\cite{xin2024bigcodec} & 80 & 1 & \texttimes & 3.25 & 0.61 & 3.59 \\
X-Codec\,2~\cite{ye2025llasa} & 50 & 1 & \checkmark & 2.63 & 0.62 & 3.68 \\
Vevo Tokenizer~\cite{zhang2025vevo} & 50 & 1 & \checkmark & 3.04 & 0.53 & 3.50 \\
CosyVoice\,2~\cite{du2024cosyvoice} & 25 & 1 & \checkmark & 4.10 & 0.68 & 3.65 \\
\bottomrule
\end{tabular}%
}
\end{table}

CosyVoice\,2's semantic tokenizer achieves the highest speaker similarity (0.68) among all compared tokenizers, with competitive WER (4.10) and UTMOS (3.65). Its low frame rate of 25\,Hz with a single codebook is particularly advantageous for autoregressive language modeling: the reduced sequence length lowers both training and inference cost, while a single codebook eliminates the architectural complexity of multi-codebook generation strategies (e.g., delayed patterns or residual prediction). The tokenizer is decoupled from our main contributions and can be replaced as better alternatives emerge.

\end{document}